\documentclass[twocolumn,prl,floatfix,superscriptaddress]{revtex4}
\usepackage[T1]{fontenc}
\usepackage{amsmath,amsfonts,amssymb,color,graphicx,graphics,latexsym,theorem,url,verbatim}
\usepackage{mathptmx}
\usepackage{bm}
\usepackage{blindtext}
\usepackage{enumitem}
\usepackage{subfigure}

\newcommand{\p}{\mathbf{p}}
\newcommand{\q}{\mathbf{q}}
\newcommand{\op}{\bm{\omega}_{\text{DP}}}
\newcommand{\os}{\bm{\omega}_{\text{DS}}}
\newcommand{\mop}{\bm{\omega}_{\text{MDP}}}
\newcommand{\mos}{\bm{\omega}_{\text{MDS}}}

\begin{document} \title{Experimental investigation of majorization uncertainty relations in the high-dimensional systems}

\author{Yuan Yuan}
\affiliation{CAS Key Laboratory of Quantum Information, University of Science and Technology of China, Hefei, 230026, China}
\affiliation{Synergetic Innovation Center of Quantum Information and Quantum Physics, University of Science and Technology of China, Hefei, Anhui 230026, China}
\author{Yunlong Xiao}
\email{yunlong.xiao@ucalgary.ca}
\affiliation{Department of Mathematics and Statistics, University of Calgary, Calgary, Alberta T2N 1N4, Canada}
\affiliation{Institute for Quantum Science and Technology, University of Calgary, Calgary, Alberta, T2N 1N4, Canada}
\author{Zhibo Hou}
\affiliation{CAS Key Laboratory of Quantum Information, University of Science and Technology of China, Hefei, 230026, China}
\affiliation{Synergetic Innovation Center of Quantum Information and Quantum Physics, University of Science and Technology of China, Hefei, Anhui 230026, China}
\author{Shao-Ming Fei}
\affiliation{School of Mathematical Sciences, Capital Normal University, Beijing 100048, China}
\affiliation{Max Planck Institute for Mathematics in the Sciences, 04103 Leipzig, Germany}
\author{Gilad~Gour}
\affiliation{Department of Mathematics and Statistics, University of Calgary, Calgary, Alberta T2N 1N4, Canada}
\affiliation{Institute for Quantum Science and Technology, University of Calgary, Calgary, Alberta, T2N 1N4, Canada}
\author{Guo-Yong Xiang}
\email{gyxiang@ustc.edu.cn}
\affiliation{CAS Key Laboratory of Quantum Information, University of Science and Technology of China, Hefei, 230026, China}
\affiliation{Synergetic Innovation Center of Quantum Information and Quantum Physics, University of Science and Technology of China, Hefei, Anhui 230026, China}
\author{Chuan-Feng Li}
\affiliation{CAS Key Laboratory of Quantum Information, University of Science and Technology of China, Hefei, 230026, China}
\affiliation{Synergetic Innovation Center of Quantum Information and Quantum Physics, University of Science and Technology of China, Hefei, Anhui 230026, China}
\author{Guang-Can Guo}
\affiliation{CAS Key Laboratory of Quantum Information, University of Science and Technology of China, Hefei, 230026, China}
\affiliation{Synergetic Innovation Center of Quantum Information and Quantum Physics, University of Science and Technology of China, Hefei, Anhui 230026, China}

\begin{abstract}
Uncertainty relation is not only of fundamental importance to quantum mechanics, but also crucial to the quantum information technology. Recently, majorization formulation of uncertainty relations (MURs) have been widely studied, ranging from two measurements to multiple measurements. Here, for the first time, we experimentally investigate MURs for two measurements and multiple measurements in the high-dimensional systems, and study the intrinsic distinction between direct-product MURs and direct-sum MURs. The experimental results reveal that by taking different nonnegative Schur-concave functions as uncertainty measure, the two types of MURs have their own particular advantages, and also verify that there exists certain case where three-measurement majorization uncertainty relation is much stronger than the one obtained by summing pairwise two-measurement uncertainty relations. Our work not only fills the gap of experimental studies of majorization uncertainty relations, but also represents an advance in quantitatively understanding and experimental verification of majorization uncertainty relations which are universal and capture the essence of uncertainty in quantum theory.
\end{abstract}

\maketitle

\textbf{\emph{Introduction.}}--Critical to almost all respects of quantum theory, including quantum information theory, is the ability to specify simultaneously the precise outcomes from incompatible measurements. This fundamental mechanism was first introduced by Heisenberg in 1927 with the name ``{\it Uncertainty Principle}'' \cite{Heisenberg1927}, which acts as one of the most striking features of quantum theory that introduces intrinsic limitations on the measurement precision of incompatible observables. At the same year, Kennard \cite{Kennard} and Weyl \cite{Weyl} derived the  first mathematical formulation for position and momentum based on standard deviation respectively. To understand the incompatibility between bounded observables, Robertson uncertainty relation was given in \cite{Robertson}.

From the information theoretic viewpoint, entropies are suitable tools in measuring the uncertainties of physical system, giving rise to entropic uncertainty relations. The first Shannon entropic uncertainty relation was given in 1983 by Deutsch \cite{Deutsch}. Later, it has been improved by Maassen and Uffink \cite{MU}.
Besides the fundamental role of entropic uncertainty relation, it also leads to a wide range of applications in certifying quantum randomness \cite{review,QR}, security proofs for quantum cryptography \cite{QC1,QC2,QC3}, detecting entanglement \cite{ED1,ED2,ED3,ED4}, Einstein-Podolsky-Rosen (EPR) steering \cite{EPRS1,EPRS2, EPRS3}, studying the wave-particle duality \cite{WP1,WP2}, and even the curved spacetime \cite{Xiao2018H}. 
Due to their importance, the uncertainty relations based on variances and entropies were tested experimentally with neutronic system \cite{neutron1,neutron2,neutron3}, photonic system \cite{photon1,photon2,photon3,photon4,photon5,photon6,photon7}, nitrogen-vacancy (NV) center \cite{NV}, and nuclear magnetic resonance (NMR) \cite{NMR}.

Only recently, physicists realize that the uncertainty of a quantum system should be nondecreasing under randomly chosen symmetry transformation and classical processing via channels \cite{Narasimhachar2016}. As such, nonnegative Schur-concave functions, including entropies, become qualified candidates for uncertainty measures and joint uncertainty relations should be characterized by majorization \cite{book, M1, M2, PRL}. Based on their formalization, majorization uncertainty relations can be divided into two categories; ``direct-product majorization uncertainty relations'' (direct-product MURs) \cite{PRL, JPA} (also known as ``universal uncertainty relations'' (UUR)) and ``direct-sum majorization uncertainty relations'' (direct-sum MURs) \cite{PRA}. The present theoretical treatment of majorization uncertainty relations (MURs) has been widely studied \cite{M5,M6,Xiao2016QM, Xiao2016U,M7}, and much of the interest in majorization uncertainty relations comes from the possibility of detecting entanglement and Einstein-Podolsky-Rosen steering \cite{Xiao2016UEF, Jia2017, Wang2018, Xiao2018Q}.


Motivated by the significant value of majorization uncertainty relations in quantum theory, we study the intrinsic distinction between direct-product MURs and direct-sum MURs, and experimentally test the two-measurement and multi-measurement MURs in the high-dimensional systems for the first time. In our experiment, we obtain a series of probability distributions induced by measuring a qudit state encoded with path and polarization degree of freedom of the photon in the eigenbasis of the observable. Furthermore, comparisons between direct-product MURs and direct-sum MURs based on different nonnegative Schur-concave functions are also demonstrated. Our work not only fills the gap of experimental studies of majorization uncertainty relations, but also enriches the experimental investigations of high-dimensional uncertainty relations and could stimulate the usages of majorization uncertainty relations in experimentally detecting entanglement and EPR steering.

\textbf{\emph{Theoretical framework.}}--We begin with a quantum state $\rho$ in a $d$-dimensional Hilbert space $\mathcal{H}$ with observables $A=\left\{|a_{j}\rangle\right\}^{d}_{j=1}$ and $B=\left\{|b_{k}\rangle\right\}^{d}_{k=1}$. Denote $p_{j}(\rho)=\langle a_{j}|\rho|a_{j}\rangle$ as the probability of receiving outcome $j\in \left\{1, 2, \ldots, d\right\}$ when performing measurement $A$, and define $\p:=(p_{j})_{j}$ as the probability distribution vector corresponds to observable $A$. Analogously, we can construct the probability distribution vector $\q:=(q_{k})_{k}$ for observable $B$. We write the direct-product between them as $\p\otimes\q$ and use $\p\oplus\q$ to present their direct-sum.

Consider now the majorization relation between vectors: a vector $\textbf{x} \in \mathbb{R}^{d}$ is \emph{majorized} by another vector $\textbf{y} \in \mathbb{R}^{d}$ whenever $\sum^{k}_{j=1}x_{j}^{\downarrow}\leqslant \sum^{k}_{j=1}y_{j}^{\downarrow}$ for all $1\leqslant k\leqslant d-1$ while $\sum^{d}_{j=1}x_{j}^{\downarrow}= \sum^{d}_{j=1}y_{j}^{\downarrow}$, and we write $\textbf{x} \prec \textbf{y}$. Here the down-arrow notation denotes that the components of the corresponding vector are ordered in decreasing order, i.e. $x_{1}^{\downarrow}\geqslant x_{2}^{\downarrow}\geqslant \cdot \cdot \cdot \geqslant x_{d}^{\downarrow}$ \cite{book}.

With the majorization relation for vectors, we now present direct-product MURs and direct-sum MURs as
\begin{align}
\p\otimes\q & \prec \op, \\
\p\oplus\q & \prec \os,
\end{align}
where $\rho$ runs over all quantum states in $\mathcal{H}$ with $\op$, $\os$ standing for the state-independent bound of direct-product MURs and direct-sum MURs respectively. Let us take any nonnegative Schur-concave function $\mathcal{U}$ to quantify the uncertainties and apply it to direct-product MURs and direct-sum MURs, which leads to
\begin{align}
\mathcal{U}(\p\otimes\q) &\geqslant \mathcal{U}( \op ),\\
\mathcal{U}(\p\oplus\q) &\geqslant \mathcal{U}( \os ).
\end{align}
The universality of MURs comes from the diversity of uncertainty measures $\mathcal{U}$ and direct-product MURs, direct-sum MURs stand for different kind of uncertainties. To construct the bound $\op$ and $\os$, we can follow the method introduced in \cite{PRL, JPA, PRA}.

We next move to describe the additivity of uncertainty measures, and call a measure $\mathcal{U}$ {\it direct-product additive} if $\mathcal{U}(\p\otimes\q)=\mathcal{U}(\p) + \mathcal{U}(\q)$. Instead of direct-product between probability distribution vectors, one can also consider direct-sum and define {\it direct-sum additive} for $\mathcal{U}$ whenever it satisfies $\mathcal{U}(\p\oplus\q)=\mathcal{U}(\p) + \mathcal{U}(\q)$. Note that the joint uncertainty $\p\oplus\q$ considered here is unnormalized and comparison between direct-product MURs and normalized direct-sum MURs is detailed in the  Appendix~\cite{SM}. Once an uncertainty measure $\mathcal{U}$ is evolved to both direct-product additive and direct-sum additive, then we call it {\it super additive} for uncertainties. It is worth to mention that $\mathcal{U}(\p\otimes\q)=\mathcal{U}(\p\oplus\q)$ whenever the uncertainty measure is super additive.
Consequently, the bound $\os$ for direct-sum MURs performs better than $\op$ in the case of super additive,
\begin{align}\label{eqs}
\mathcal{U}(\p\otimes\q)=\mathcal{U}(\p\oplus\q) \geqslant \mathcal{U}( \os ) \geqslant \mathcal{U}( \op ),
\end{align}
since $\os \prec \op$ \cite{PRA}. We remark that the well known Shannon entropy is super additive and only by applying super additive functions \cite{SM}, like Shannon entropy, direct-product MURs and direct-sum MURs are comparable. It should also be clear that direct-product MURs and direct-sum MURs have been employed to describe different type of uncertainties. For an uncertainty measure $\mathcal{U}$, in general, it can be checked that $\mathcal{U}(\p\otimes\q) \neq \mathcal{U}(\p\oplus\q)$ and hence it is meaningless to state that direct-sum MURs performs better than direct-product MURs and vice versa.

One of the main goals in the study of uncertainty relations is the quantification of the joint uncertainty of incompatible observables. Direct-product MURs and direct-sum MURs provide us two different methods to quantify joint uncertainty between incompatible observables. Relations between direct-product MURs and direct-sum MURs are of fundamental importance both for the theoretical characterization of joint uncertainties, as well as the experimental implementation. Quite uncannily, we find that for some eligible uncertainty measure $\mathcal{U}$, direct-product MURs and direct-sum MURs are given by
\begin{align}\label{eq1}
\mathcal{U}(\p\oplus\q) \geqslant \mathcal{U}( \os ) > \mathcal{U}(\p\otimes\q) \geqslant \mathcal{U}( \op ),
\end{align}
for some quantum state $\rho$.

\begin{figure*}[tbph]
\includegraphics [width=16cm,height=8cm]{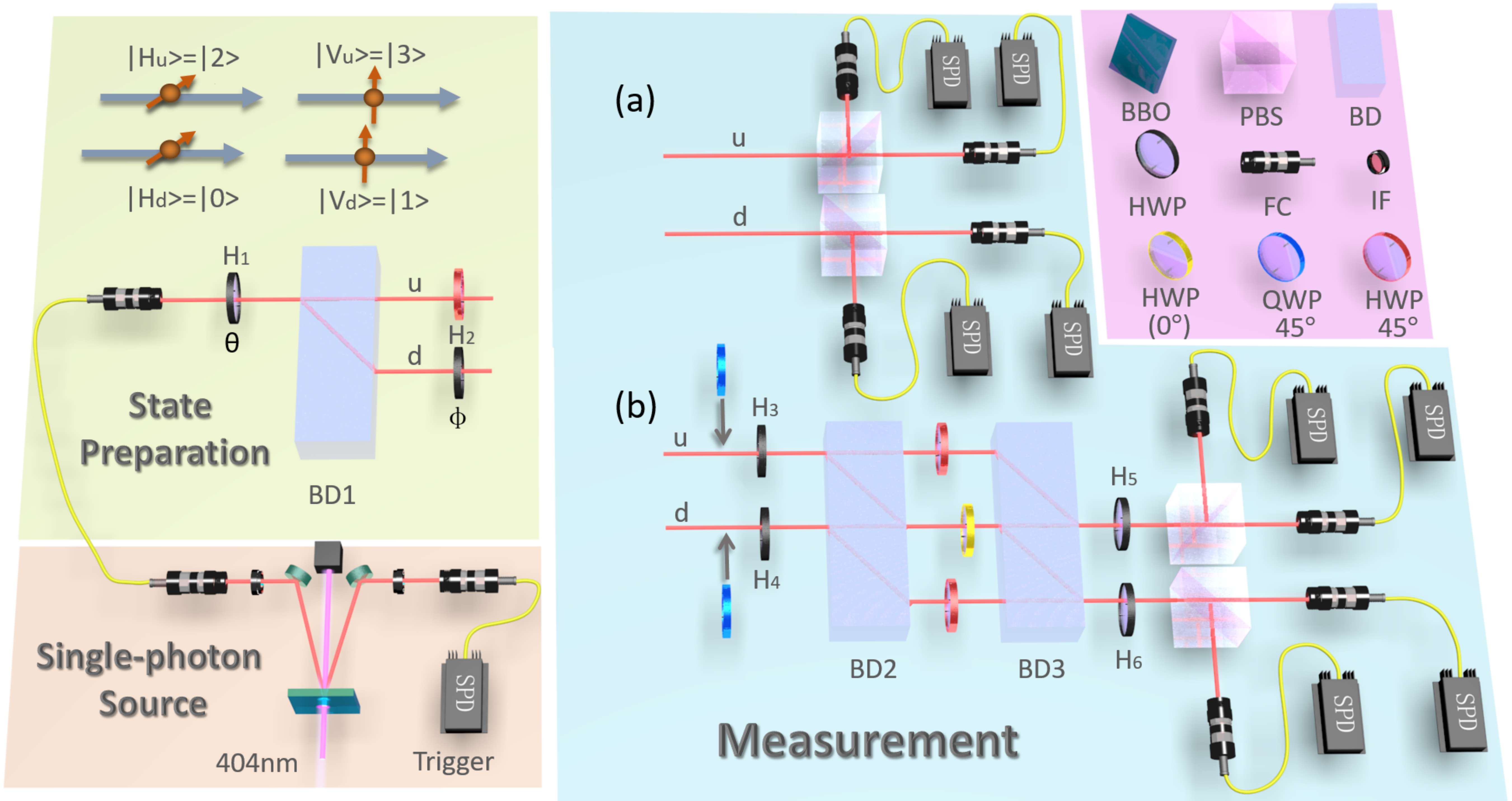}
\caption{Experimental setup. In the single-photon source module, the photon pairs generated in spontaneous parametric down-conversion are coupled into single-mode fibers separately. One photon is detected by a single-photon detector (SPD) acting as a trigger. In the state preparation module, a qudit state is encoded by four modes of the single photons. H and V denote the horizontal polarization and vertical polarization of the photon, respectively. The subscripts u and d represent the upper and lower spatial modes of the photon, respectively. The half-wave plates (H$_{1}$, H$_{2}$) and beam displacer (BD1) are used to generate desired qudit state. In the measurement module, the red HWPs with an angle of $45^{\circ}$ and beam displacers (BDs) comprise the interferometric network to perform the desired measurement; the yellow HWP with an angle of $0^{\circ}$ are inserted into the middle path to compensate the optical path difference between the upper and lower spatial modes. To realize measurement $B$ shown in Eq. (\ref{2measurement}), two quarter-wave plates are need to be inserted in device (b). Output photons are detected by single-photon detectors (SPDs).}
\label{fig:setup}
\end{figure*}

Let us now construct such uncertainty measure $\mathcal{U}$. First define the summation function $\mathcal{S}$ as $\mathcal{S}(\mathbf{u}):=\sum_{l}u_{l}=\left\lVert \mathbf{u} \right\rVert_{1}$ with $\mathbf{u}=(u_{1}, u_{2}, \ldots, u_{d})$. Another important function $\mathcal{M}$ is defined as $\mathcal{M}(\mathbf{u}):=\max_{l}u_{l}=2^{-H_{\text{min}}(\mathbf{u})}$. And hence it is easy to check that $\mathcal{U}:=\mathcal{S}-\mathcal{M}$ is a nonnegative Schur-concave functions; take two vectors satisfying $x \prec y$, and based on the definition of $\mathcal{U}$ we have $\mathcal{U}(x)=\sum^{d}_{j=2}x_{j}^{\downarrow}\geqslant \sum^{d}_{j=2}y_{j}^{\downarrow}=\mathcal{U}(y)$. Specifically this function, which combines $\mathcal{S}$ and $\mathcal{M}$ together, is a qualified uncertainty measure and satisfies Eq. (\ref{eq1}) for some quantum states and measurements. Moreover, specific examples are given in the following experimental demonstration.

In principle, direct-product MURs and direct-sum MURs do not have to be comparable and their joint uncertainty can be quantified by their bound. However, we can compare their differences by checking which bound approximates their joint uncertainty better since joint uncertainties are often classified by their bounds.
Take any nonnegative Schur-concave function $\mathcal{U}$, which leads to two nonnegative quantities $\xi_{DS}:=\mathcal{U}(\p\oplus\q) - \mathcal{U}( \os )$ and $\xi_{DP}:=\mathcal{U}(\p\otimes\q) - \mathcal{U}( \op )$. To determine whether the bound $\op$ approximates direct-product MURs better than $\os$ approximates direct-sum MURs, we simply compare the numerical value of $\xi_{DS}$ and $\xi_{DP}$. And how such bounds contribute to the joint uncertainties are depicted in our experiment.

The question arises whether the majorization uncertainty relations, including direct-product MURs and direct-sum MURs, can be extended to multi-measurement cases and the answer is clear. We assume that the experimenter performs measurements $C_{l}$ ($l>2$) to the quantum state $\rho$ and denote their probability distributions as $\p_{l}$, then the multi-measurement joint uncertainties can be quantified by
\begin{align}
\bigotimes_{l} \p_{l} &\prec \mop,\notag\\
\bigoplus_{l} \p_{l} &\prec \mos,
\end{align}
where $\mop$ and $\mos$ are the bounds for multi-measurement direct-product MURs and direct-sum MURs, constructed in \cite{PRL} and \cite{PRA} respectively.

To experimentally test direct-product MURs, direct-sum MURs and their relations, we choose a family of $4$-dimensional states
\begin{equation} \label{state}
\begin{aligned}
|\psi_{\theta,\phi}\rangle&=\cos\theta\sin\phi|0\rangle+\cos\theta\cos\phi|1\rangle+\sin\theta|2\rangle
\\&=(\cos\theta\sin\phi,\cos\theta\cos\phi,\sin\theta,0)^{\top},
\end{aligned}
\end{equation}
where $|0\rangle, |1\rangle, |2\rangle, |3\rangle$ are orthonormal basis states in a $4$-dimensional Hilbert space. Correspondingly, we measure the state Eq. (\ref{state}) in the following measurements, including two measurements
\begin{equation}\label{2measurement}
\begin{aligned}
&A=\{|0\rangle, |1\rangle, |2\rangle, |3\rangle\}
\\&B=\begin{aligned}&\{\frac{|0\rangle-i|1\rangle-i|2\rangle+|3\rangle}{2}, \frac{|0\rangle-i|1\rangle+i|2\rangle-|3\rangle}{2}, \\&\frac{|0\rangle+i|1\rangle-i|2\rangle-|3\rangle}{2},\frac{|0\rangle+i|1\rangle+i|2\rangle+|3\rangle}{2}\}\end{aligned}
\end{aligned}
\end{equation}
and three measurements
\begin{equation}\label{3measurement}
\begin{aligned}
&C_{1}=\{|0\rangle, |1\rangle, |2\rangle, |3\rangle\}
\\&C_{2}=\{|0\rangle, \frac{|2\rangle+|3\rangle}{\sqrt{2}}, \frac{|1\rangle+|2\rangle-|3\rangle}{\sqrt{3}}, \frac{2|1\rangle-|2\rangle+|3\rangle}{\sqrt{6}}\}
\\&C_{3}=\{\frac{|2\rangle+|3\rangle}{\sqrt{2}}, |1\rangle, \frac{|0\rangle+|2\rangle-|3\rangle}{\sqrt{3}}, \frac{2|0\rangle-|2\rangle+|3\rangle}{\sqrt{6}}\}.
\end{aligned}
\end{equation}

\begin{figure}[tbph]
\includegraphics [width=8.7cm,height=7.5cm]{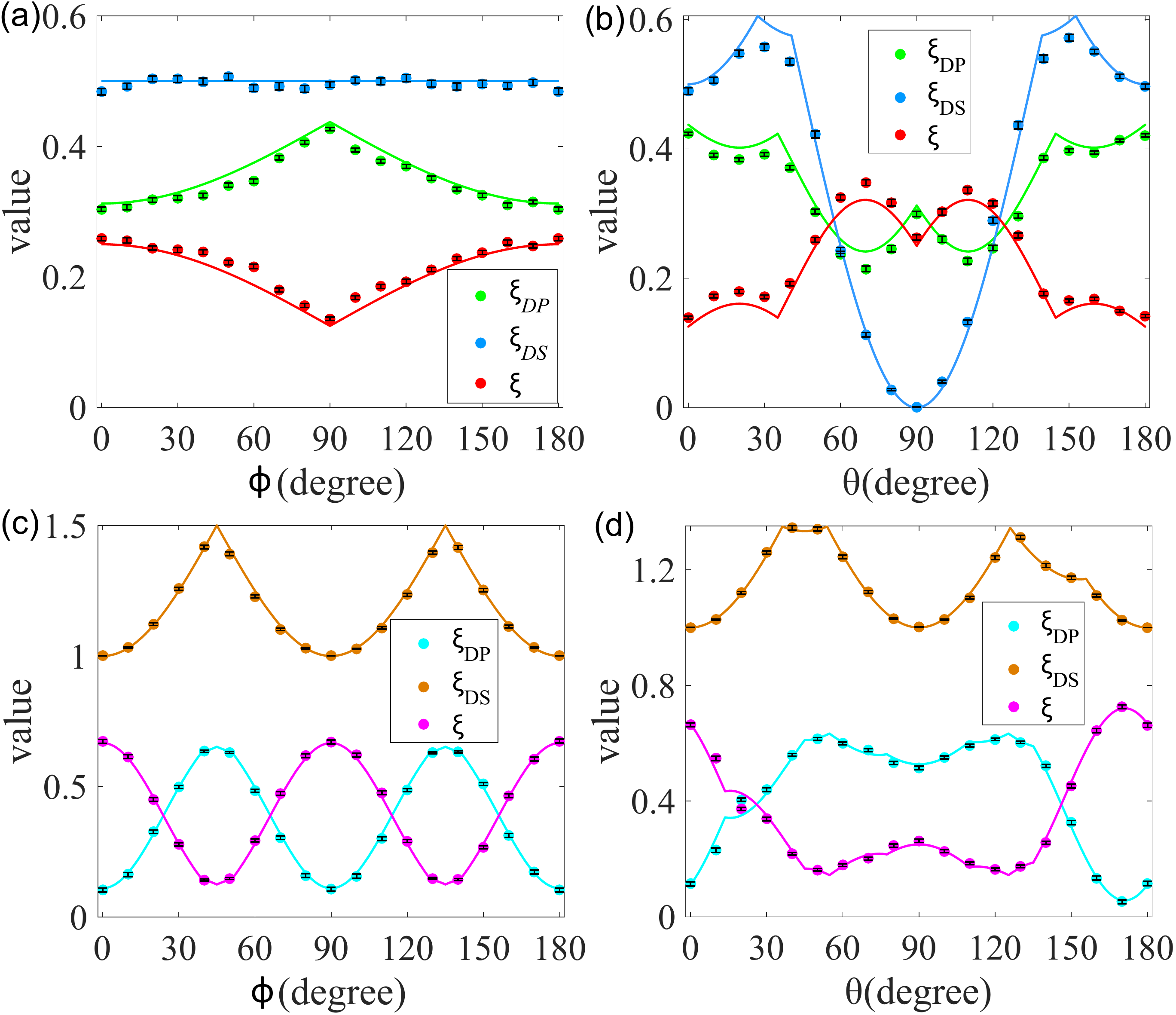}
\caption{Experimental results for comparing the differences of direct-product MURs and direct-sum MURs by checking which bound approximates their joint uncertainty better. The curve is theoretical value, and the dot is experimental value. The three quantities are expressed as $\xi_{DS}:=\mathcal{U}(\p\oplus\q) - \mathcal{U}( \os )$, $\xi_{DP}:=\mathcal{U}(\p\otimes\q) - \mathcal{U}( \op )$ and $\xi=\mathcal{U}( \os )- \mathcal{U}(\p\otimes\q)$ with $\mathcal{U}:=\mathcal{S}-\mathcal{M}$. Fig. 2(a) and (b) show the experimental results by measuring states $|\psi_{\pi/4,\phi}\rangle$ and $|\psi_{\theta,\pi/4}\rangle$ in the two measurements Eq. (\ref{2measurement}), respectively. Fig. 2(c) and (d) show the experimental results by measuring states $|\psi_{\pi,\phi}\rangle$ and $|\psi_{\theta,\pi/2}\rangle$ in the three measurements Eq. (\ref{3measurement}), respectively. The error bar
denotes the standard deviation.}
\label{fig:second}
\end{figure}

\textbf{\emph{Experimental demonstration.}}--We experimentally test direct-product MURs, direct-sum MURs and their relations with a linear optical system. The experimental setup, illustrated in Fig. 1, consists of three modules, including the single-photon source module (see Appendix C for details), the state preparation module, and the measurement module. In the state preparation module, a qudit state is encoded by four modes of the heralded single photon, and the basis states $|0\rangle$, $|1\rangle$, $|2\rangle$, and $|3\rangle$ are encoded by the horizontal polarization of the photon in the lower mode, the vertical polarization of the photon in the lower mode, the horizontal polarization of the photon in the upper mode, and the vertical polarization of the photon in the upper mode, respectively. The beam displacer (BD) causes the vertical polarized photons to be transmitted directly and the horizontal polarized photons to undergo a $4$-mm lateral displacement, so the photon passes through a half-wave plate (H$_{1}$) with a certain setting angle and then is split by BD1 into two parallel spatial modes--upper and lower modes. H$_{1}$ and H$_{2}$ control the parameters $\theta$ and $\phi$, respectively, thus the photon is prepared in the desired state $|\psi_{\theta,\phi}\rangle$. In the measurement module, measuring device (a) is used to realize measurement $A$ and measurement $C_{1}$. When the measuring device (b) has quarter-wave plates with an angle of $45^{\circ}$, it is used to realize measurement $B$, and the setting angles of H$_{3}$--H$_{6}$ are $45^{\circ}$, $0^{\circ}$, $22.5^{\circ}$, and $22.5^{\circ}$, respectively. The measuring device (b) without quarter-wave plates is used to realize measurement $C_{2}(C_{3})$ when the setting angles of H$_{3}$--H$_{6}$ are $22.5^{\circ}$, $0^{\circ}(45^{\circ})$, $27.4^{\circ}$, and $0^{\circ}$, respectively.

\begin{figure}[tbph]
\includegraphics [width=8.8cm,height=8.4cm]{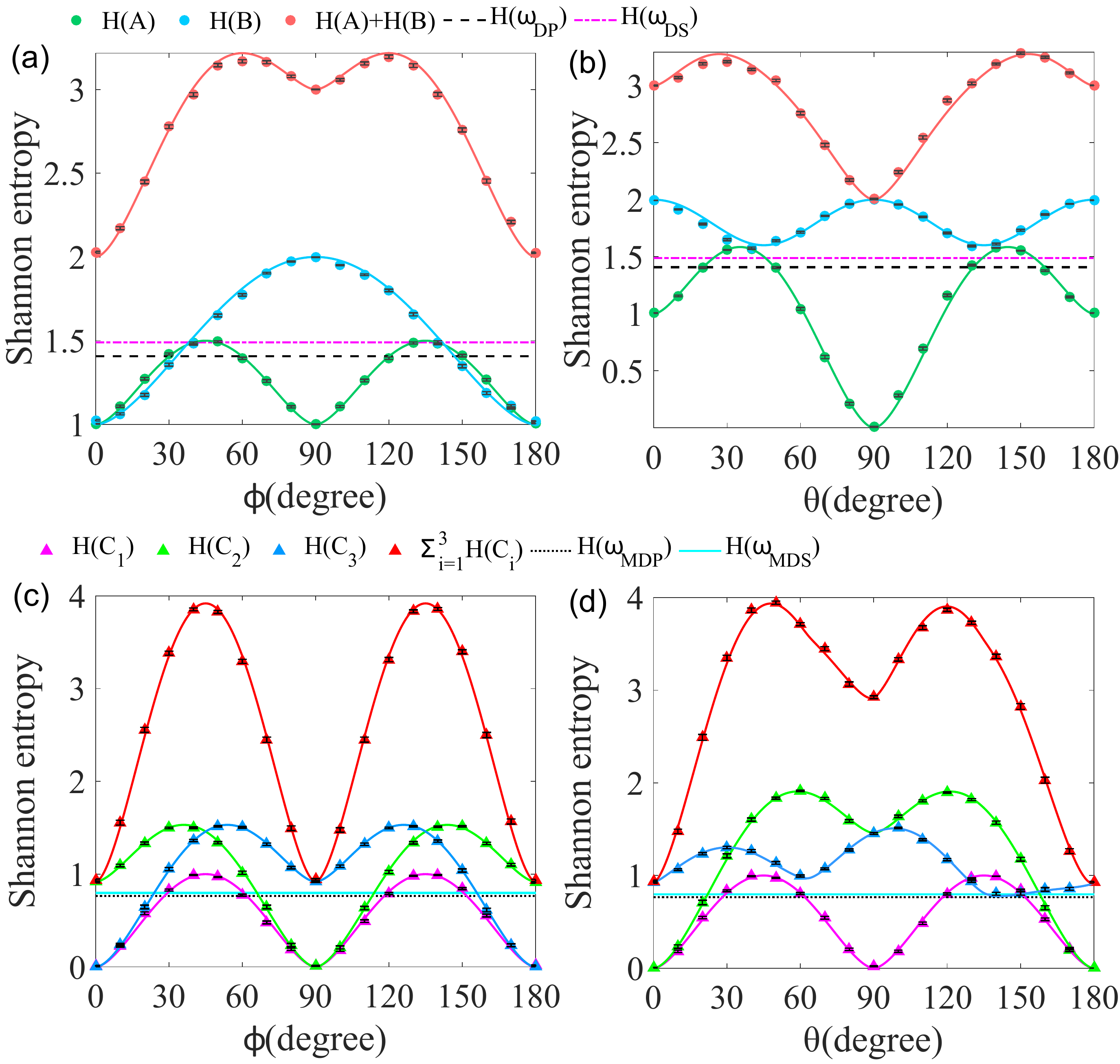}
\caption{Experimental results for majorization uncertainty relations based on Shannon entropy measure. Fig. 3(a) and (b) show the results for measuring states $|\psi_{\pi/4,\phi}\rangle$ and $|\psi_{\theta,\pi/4}\rangle$ in the measurements $A$ and $B$, respectively. The green dots and curve represent Shannon entropy of the probability distributions induced by measurement $A$. The blue dots and curve represent Shannon entropy of the probability distributions induced by measurement $B$. The red dots and curve represent the sum of $H(A)$ and $H(B)$. The dashed line and dash-dotted line are the bounds of the two-measurement direct-product MURs and direct-sum MURs, respectively. Fig. 3(c) and (d) show the results for measuring states $|\psi_{\pi,\phi}\rangle$ and $|\psi_{\theta,\pi/2}\rangle$ in the measurements $C_{1}$, $C_{2}$ and $C_{3}$, respectively.  The magenta, green, blue triangles and curve represent Shannon entropy of the probability distributions obtained by measuring the $C_{1}$, $C_{2}$ and $C_{3}$, respectively. The red triangles and curve represent the sum of the $H(C_{1})$, $H(C_{2})$ and $H(C_{3})$. The dotted line and solid line are the bounds of the three-measurement direct-product MURs and direct-sum MURs, respectively. The error bar denotes the standard deviation.}
\label{fig:first}
\end{figure}

\textbf{\emph{Experimental results.}}--To test the relation Eq. (\ref{eq1}) and compare the differences of direct-product MURs and direct-sum MURs by checking which bound approximates their joint uncertainty better, we experimentally obtain a series of probability distributions induced by measuring state $|\psi_{\theta,\phi}\rangle$ in the measurements Eq. (\ref{2measurement}) and Eq. (\ref{3measurement}). Under Schur-concave function $\mathcal{U}:=\mathcal{S}-\mathcal{M}$, Fig. 2 depicts the experimental values of $\xi_{DP}$, $\xi_{DS}$ and $\xi$ with $\xi=\mathcal{U}( \os )- \mathcal{U}(\p\otimes\q)$. We remark that $\xi$ stands for the difference between the bounds of direct-sum MURs $\mathcal{U}( \os )$ and the joint uncertainty of direct-product MURs $\mathcal{U}(\p\otimes\q)$. Whenever the quantity $\xi$ is positive, we cannot use $\mathcal{U}( \os )$ to bound $\mathcal{U}(\p\otimes\q)$, and thus cannot state that the bound of direct-sum MURs outperforms the bound of direct-product MURs. Fig. 2(a) and (b) show the experimental results by measuring states $|\psi_{\pi/4,\phi}\rangle$ and $|\psi_{\theta,\pi/4}\rangle$ in the two measurements Eq. (\ref{2measurement}), respectively. In Fig. 2(a), $\xi_{DP}$ is less than $\xi_{DS}$, which implies that the bound of direct-product MURs approximates their joint uncertainty better, but in Fig. 2(b) which bound is better depends on the state. Fig. 2(c) and Fig. 2(d) demonstrate the experimental results by measuring states $|\psi_{\pi,\phi}\rangle$ and $|\psi_{\theta,\pi/2}\rangle$ in the three measurements Eq. (\ref{3measurement}), respectively. We can see that $\xi_{DP}$ is always less than $\xi_{DS}$ with the variation of the state $|\psi_{\theta,\phi}\rangle$, which means that the bound of direct-product MURs approximates their joint uncertainty better in the case of these measurements and states than that of direct-sum MURs. From Fig. 2, it is clear that $\xi$ is always greater than zero, and hence both the mathematical inequality Eq. (\ref{eq1}) and its physical significance have been verified.

From the information theoretic viewpoint, entropic functions are important nonnegative Schur-concave functions. We are interested in entropic uncertainty relations, which can be derived by setting $\mathcal{U}$ as the Shannon entropy $H$, and can be expressed as $H(\bigotimes_{l} \p_{l}) \geqslant H(\mop)$ and $H(\bigoplus_{l} \p_{l}) \geqslant H(\mos)$. Here we also compare direct-product MURs and direct-sum MURs with Shannon entropy and the experimental results are portrayed in Fig. 3. Fig. 3(a) and Fig. 3(b) exhibit the experimental results for two-measurement majorization uncertainty relation based on Shannon entropy measure; by choosing measurements $A$ and $B$, the bound of direct-product MURs $\op$ is given by $\op=(0.5625, 0.1661, 0.2714)$ which implies $H(\op)=1.4077$ shown by the dashed line in Fig. 3(a) and Fig. 3(b). On the other hand, $\os = (0.5, 0.2071, 0.2929)$, and as a consequence we have $H(\os)=1.4893$ shown by the dash-dotted line in Fig. 3(a) and Fig. 3(b). Now we are in the position in describing the experimental results for three-measurement majorization uncertainty relation based on Shannon entropy measure. The bound of direct-product MURs, $\mop$, is given by $(0.7773, 0.2227)$, which leads to $H(\mop)=0.7651$, as shown by the dotted line in the Fig. 3(c) and Fig. 3(d). Meanwhile $\mos$ is given by $(1, 1, 0.7583, 0.2417)$, with $H(\mos)=0.7979$, as shown by the line in the Fig. 3(c) and Fig. 3(d). For all these cases with two measurements or three measurements, the bound of direct-sum MURs is always tighter than the bound of direct-product MURs, which verified the relation Eq. (\ref{eqs}). Finally, we remark that general multi-measurement uncertainty relation can be generated by summing pairwise two-measurement uncertainty relations trivially, while in our three-measurement case, pairwise two-measurement uncertainty relations lead to a trivial bound of zero, but the majorization uncertainty relation for the three measurements remains nontrivial.

\textbf{\emph{Conclusions.}}--To summarize, we have experimentally studied the high-dimensional majorization uncertainty relations, namely direct-product MURs and direct-sum MURs, with a linear optical system and analyzed the essential difference between them for the first time. Direct-product MURs and directsum MURs are different types of joint uncertainty, taking different nonnegative Schur-concave functions as uncertainty measure, these two majorization uncertainty relations have their own particular advantages. From experimental data, direct-sum MURs are stronger than direct-product MURs under Shannon entropy. Moreover, we have also experimentally verified that there exists certain case where three-measurement majorization uncertainty relation is much stronger (with nontrivial bound) than the one obtained by summing pairwise two-measurement uncertainty relations (with trivial bound). We expect that our theoretical analysis and experimental tests will lead to further results that enrich the studies of high-dimensional uncertainty relations. Specifically, our work could provide potential applications in the field of quantum information technology, such as quantum precision measurement, detecting entanglement and EPR steering.

{\bf Acknowledgements:}
This work is supported by the National Natural Science Foundation of China (Grants No. 11574291 and No. 11774334), China Postdoctoral Science Foundation (Grant No. 2016M602012 and No. 2018T110618), National Key Research and Development Program of China (Grants No. 2016YFA0301700 and No.2017YFA0304100), and Anhui Initiative in Quantum Information Technologies. Y. Xiao and G. Gour acknowledge NSERC support. S.-M. Fei acknowledges financial
support from the National Natural Science Foundation of China under Grant No. 11675113 and Beijing Municipal Commission of Education (KZ201810028042).

\appendix*
\setcounter{equation}{0}
\subsection*{Appendix A: Normalized direct-sum MURs}

\begin{figure}[tbph]
\begin{center}
\includegraphics[width=8.8cm,height=6.8cm]{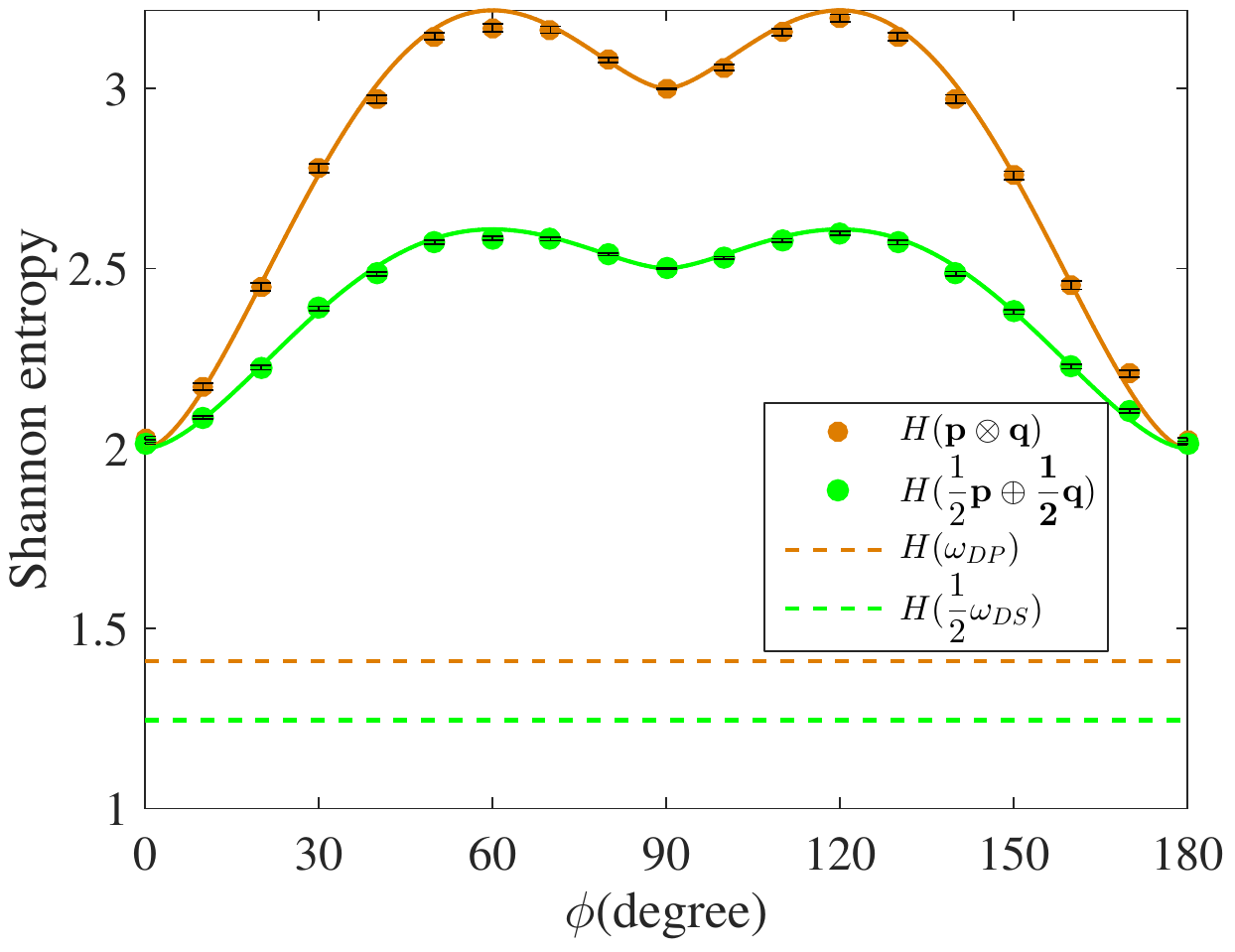} \\
\includegraphics[width=8.8cm,height=6.8cm]{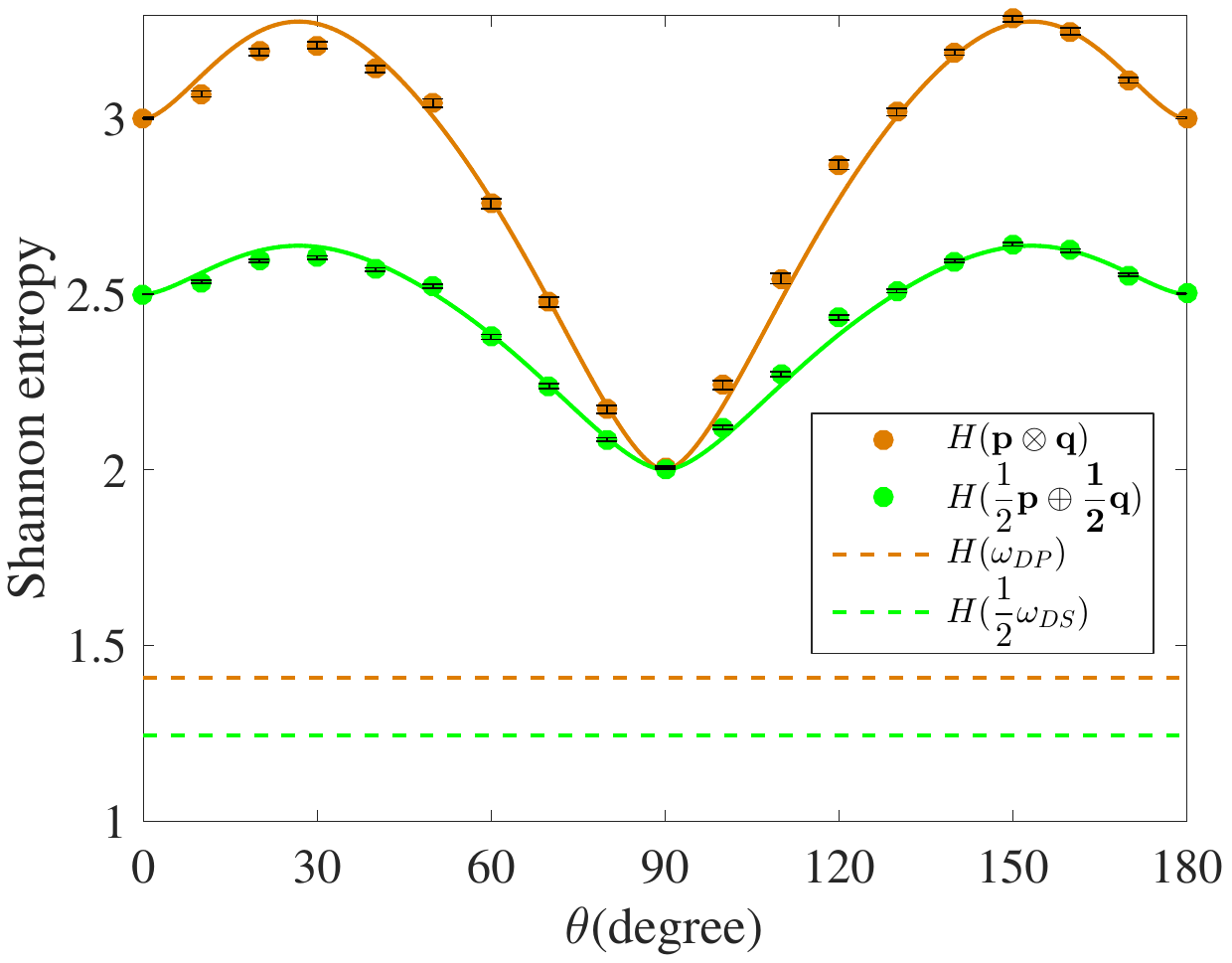}
\end{center}
\caption{Experimental results for comparison between normalized direct-sum MURs and direct-product MURs based on Shannon entropy measure. The two pictures show the results for measuring states $|\psi_{\pi/4,\phi}\rangle$ and $|\psi_{\theta,\pi/4}\rangle$ in the observables $A$ and $B$, respectively.}
\label{sm1}
\end{figure}

The direct-sum MURs were first given in \cite{PRA} with the form
\begin{align}
\p\oplus\q \prec \os,
\end{align}
for probability distributions $\p$ and $\q$. However, unlike $\p\otimes\q$ constructed in direct-product MURs \cite{PRL}, $\p\oplus\q$ is not a probability distribution. In order to derive a normalized direct-sum MURs, we simply take the weight $1/2$
\begin{align}
\frac{1}{2}\p\oplus\frac{1}{2}\q & \prec \frac{1}{2}\os.
\end{align}
And now we compare the normalized direct-sum MURs $\frac{1}{2}\p\oplus\frac{1}{2}\q \prec \frac{1}{2}\os$ with direct-product MURs $\p\otimes\q \prec \op$; by taking the states shown in the main text
\begin{equation}
\begin{aligned}
|\psi_{\theta,\phi}\rangle&=\cos\theta\sin\phi|0\rangle+\cos\theta\cos\phi|1\rangle+\sin\theta|2\rangle
\\&=(\cos\theta\sin\phi,\cos\theta\cos\phi,\sin\theta,0)^{\top},
\end{aligned}
\end{equation}
and measurements $A$, $B$ with the following eigenvectors
\begin{equation}
\begin{aligned}
&A=\{|0\rangle, |1\rangle, |2\rangle, |3\rangle\}
\\&B=\begin{aligned}&\{\frac{|0\rangle-i|1\rangle-i|2\rangle+|3\rangle}{2}, \frac{|0\rangle-i|1\rangle+i|2\rangle-|3\rangle}{2}, \\&\frac{|0\rangle+i|1\rangle-i|2\rangle-|3\rangle}{2},\frac{|0\rangle+i|1\rangle+i|2\rangle+|3\rangle}{2}\}.\end{aligned}
\end{aligned}
\end{equation}
We depicted the pictures of $H\left(\p\otimes\q\right)$, $H\left(\frac{1}{2}\p\oplus\frac{1}{2}\q\right)$, $H\left(\op\right)$, and $H\left(\frac{1}{2}\os\right)$ in Fig. \ref{sm1}.

\subsection*{Appendix B: Super Additivity}

To be comparable for direct-product MURs and direct-sum MURs, we should choose a uncertainty measure $\mathcal{U}$ that are both Schur-concave and super additive. Clearly Shannon entropy is a qualified candidate. The question, thus, naturally arises: is there another function satisfies the following properties:
\begin{description}[align=left]
\item [Property 1.] $\mathcal{U}$ should be continuous in $\p$ and $\q$.
\item [Property 2.] $\mathcal{U}$ should be a Schur-concave function.
\item [Property 3.] $\mathcal{U}$ should be super additive, i.e.
	\begin{align}
        \mathcal{U}(\p\otimes\q)&=\mathcal{U}(\p) + \mathcal{U}(\q),\\
        \mathcal{U}(\p\oplus\q)&=\mathcal{U}(\p) + \mathcal{U}(\q).
        \end{align}
\end{description}
Or will these properties lead to a unique function (up to a scalar)? Since we can take $\q$ as $\left(1, 0, \ldots, 0\right)$, and then $\mathcal{U}(\p\otimes\q)= \mathcal{U}(\p)$ which is continuous in the $p_{i}$ while $\p=\left(p_{i}\right)_{i}$. Moreover, due to the Schur-concavity, $\mathcal{U}$ is a monotonic increasing function of $d$ when taking $p_{i}=\frac{1}{d}$. In addition, if $\mathcal{U}$ complies with the composition law for compound experiments, then there is only one possible expression for $\mathcal{U}$, i.e. Shannon entropy (up to a scalar). Namely, if there is a measure, say $\mathcal{U}(\p)=\mathcal{U}\left(p_{1}, p_{2}, \ldots, p_{d}\right)$ which is required to meet the following three properties:
\begin{description}[align=left]
\item [Property 4.] $\mathcal{U}$ should be continuous in $\p$.
\item [Property 5.] If all the $p_{i}$ are equal, $p_{i}=\frac{1}{d}$, then $\mathcal{U}$ should be a monotonic increasing function of $d$. With equally $d$ likely events there is more choice, or uncertainty, when there are more possible events.
\item [Property 6. (Composition Law)] If a choice be broken down into two successive choices, the original $\mathcal{U}$ should be the weighted sum of the individual values of $\mathcal{U}$.
\end{description}
Then the only $\mathcal{U}$ satisfying the three above assumptions is of the form \cite{Shannon1948}:
\begin{align}
\mathcal{U}(\p)=k\cdot\left(-\sum\limits_{i=1}^{d}p_{i}\log p_{i}\right),
\end{align}
where $k$ is a positive constant. Whenever a function $\mathcal{U}$ satisfies Property 1 and Property 2, it will meet Property 4 and Property 5 automatically. However, super additivity differs with the Composition Law, and this leads to function satisfied Property 1, 2, and 3 other than Shannon entropy.

For example, consider the composition between logarithmic function and elementary symmetric function:
\begin{align}
\mathcal{F}(\p):=\log\left(\prod\limits_{i=1}^{d}p_{i}\right).
\end{align}
Here $\mathcal{F}$ satisfies Properties 1, 2, and the direct-product MURs is read as
\begin{align}
\mathcal{F}(\p\otimes\q)&=\log\left(\prod\limits_{i, j}p_{i}q_{j}\right)\notag\\
&=\log\left(\prod\limits_{i}p_{i}\cdot\prod\limits_{j}q_{j}\right)\notag\\
&=\log\left(\prod\limits_{i}p_{i}\right)+\log\left(\prod\limits_{j}q_{j}\right)\notag\\
&=\mathcal{F}(\p)+\mathcal{F}(\q),
\end{align}
where the probability distributions $\p$ and $\q$ are defined as $\left(p_{i}\right)_{i}$ and $\left(q_{j}\right)_{j}$. On the other hand, direct-sum MURs is written as
\begin{align}
\mathcal{F}(\p\oplus\q)&=\log\left(\prod\limits_{i, j}p_{i}q_{j}\right)=\mathcal{F}(\p\otimes\q),
\end{align}
hence, $\mathcal{F}$ meets Property 3. To summarize, we derive a function $\mathcal{F}$, which is valid for Properties 1, 2, and 3. However $\mathcal{F}$ is not a good uncertainty measure, since $\mathcal{F}(\os)$ and $\mathcal{F}(\op)$ are not well defined (due to $\log0$). Whether there exists another function that obeys Property 1, 2, and 3 is an open question, and one may conjecture that $\mathcal{F}$, Shannon entropy $H$ and the convex combinations of $\mathcal{F}$ and $H$ are the only suitable candidates.

\subsection*{Appendix C: Details for the single-photon source module of the experimental setup}
In the single-photon source module, a 80-mW cw laser with a 404-nm wavelength (linewidth=5 MHz) pumps a type-II beamlike phase-matching beta-barium-borate (BBO, 6.0$\times$6.0$\times$2.0 ~mm$^{3}$, $\theta=40.98^{\circ}$) crystal to produce a pair of photons with wavelength $\lambda = 808$~nm. After being redirected by mirrors and passed through the interference filters (IF, $\bigtriangleup\lambda=3$~nm, $\lambda=808$~nm), the photon pairs generated in spontaneous parametric down-conversion are coupled into single-mode fibers separately. One photon is detected by a single-photon detector acting as a trigger. The total coincidence counts are approximately $4\times10^{3}$ per second.


\begin{thebibliography}{99}
\bibitem{SM} See Appendix for details of the comparison between direct-product MURs
$\p\otimes\q \prec \op$ and normalized direct-sum MURs $\frac{1}{2}\p\oplus\frac{1}{2}\q  \prec \frac{1}{2}\os$, and discussions on super additive functions; other than Shannon entropy, another Schur-concave and super additive function has been constructed in  Appendix B.
\end{thebibliography}
\end{document}